\documentclass[english,aps,preprint,showpacs]{revtex4}
\usepackage{subfigure}
\usepackage{graphicx,dcolumn,bm,color,latexsym,amssymb}
\usepackage[latin1]{inputenc}
\usepackage{multirow}
\definecolor{Blue}{rgb}{0.0,0.0,1}
\definecolor{Red}{rgb}{1,0.0,0.0}
\definecolor{Green}{rgb}{0,0.5,0.0}

\begin{document}

\title{Quantum state tomography and quantum logical operations in a three qubits NMR quadrupolar system}

\author{A. G. Araujo-Ferreira}
 \email{avatar@ifsc.usp.br}
\affiliation{Instituto de F\'{\i}sica de S\~{a}o Carlos, Universidade de S\~{a}o
Paulo, P.O. Box 369, 13560-970 S\~{a}o Carlos, SP, Brazil}
\author{C. A. Brasil}
 \email{carlosbrasil@ifsc.usp.br}
\affiliation{Instituto de F\'{\i}sica de S\~{a}o Carlos, Universidade de S\~{a}o
Paulo, P.O. Box 369, 13560-970 S\~{a}o Carlos, SP, Brazil}
\author{J. Teles}
\affiliation{Centro de Ci\^{e}ncias Agr\'{a}rias, Universidade Federal de S\~{a}o Carlos, Rodovia Anhanguera SP-300, km 174, 13600-970, Araras, SP, Brazil}
\author{D. O. Soares-Pinto}
\affiliation{Instituto de F\'{\i}sica de S\~{a}o Carlos, Universidade de S\~{a}o
Paulo, P.O. Box 369, 13560-970 S\~{a}o Carlos, SP, Brazil}
\author{E. R. deAzevedo}
\affiliation{Instituto de F\'{\i}sica de S\~{a}o Carlos, Universidade de S\~{a}o
Paulo, P.O. Box 369, 13560-970 S\~{a}o Carlos, SP, Brazil}
\author{T. J. Bonagamba}
\affiliation{Instituto de F\'{\i}sica de S\~{a}o Carlos, Universidade de S\~{a}o
Paulo, P.O. Box 369, 13560-970 S\~{a}o Carlos, SP, Brazil}
\date{\today}

\pacs{03.65.Ud, 03.30.+p, 03.67.Bg, 04.62.+v}

\date{\today}

\begin{abstract}
\noindent In this work, we present an implementation of quantum logic gates and algorithms in a three effective qubits system, represented by a ($I = 7/2$) NMR quadrupolar nuclei. To implement these protocols we have used the strong modulating pulses (SMP). The various stages of each implementation were verified by quantum state tomography (QST). It is presented here the results for the computational  base states, Toffolli logic gates, and Deutsch-Jozsa and Grover algorithms. Also, we discuss the difficulties and advantages of implementing such protocols using the SMP technique in quadrupolar systems.
\end{abstract}

\pacs{03.67.Ac,03.67.Lx,82.56-b,82.56.Jn}

\maketitle


\section{Introduction}

The characterization of the state of a quantum system is one of the most important steps in quantum physics and, in particular, in quantum computing (QC) \cite{ref1,ref2,ref3,ref4,ref5}. Nuclear magnetic resonance (NMR) has provided unique methods for demonstrating the realization of quantum logical operations and characterizing the quantum state of spin systems. The great sucess achieved so far by NMR QC is related to its ability for controlling the dynamics of the spins through radio frequency (RF) pulses \cite{livro,relax2}. This great control also enables the creation of pulse sequences for the determination of elements of the nuclear spins density matrix, called the Quantum State Tomography process (QST). The first NMR QST method was developed by Chuang \emph{et al.} \cite{CHUANG} and optimized by Long \emph{et al.} \cite{Hong} for systems of heteronuclear coupled spins $1/2$. It consists basically in applying a set of specially designed rotations on the different spins and reconstructing the density matrix from the resulting NMR spectra. This method was later adapted for homonuclear coupled spins $1/2$ \cite{CHUANG2,daskumar} and also for quadrupolar spin $3/2$ systems \cite{KAMP, Khitrin,sinha, bonkjmr,bonkpra,bulnes}, where nonselective pulses were replaced by transition-selective RF pulses. The disadvantage of such method is the long time required to perform a selective pulse, producing considerable unwanted evolution of the system during the QST process. Besides, the mathematical formalism used to manipulate the transition operators \cite{vega1,vega2,ernst} are quite complicated and the resulting equations are not easily generalizable for nuclei with nuclear spin greater than $3/2$. Thus, a new method for tomography \cite{telesartigo}, based on nonselective pulses \cite{SMPKAMP}, was proposed and generalized to any spin system.

Previous works, that used nonselective pulses to implement QST, explored a spin $3/2$ quadrupolar system (an effective two-qubits system) showing the great acurracy of the technique \cite{telesartigo, KAMP} also for relaxation studies \cite{relax1,relax3,relax5,relax4} and simulation of quantum systems \cite{bose} . However, when the spin quantum number is increased, some technical problems during the implementation may appear. For example, during each pulse of the QST process, several experimental errors, like oscillating field inhomogeneities, pulse imperfections, quadrupolar evolution, among others, may compromise the method performance.

We report the QST adaptation for spin $I=7/2$ in Section II. Testing procedures, results that confirm its efficiency  and simulations that took experimental limitations into account are shown in Section III. In Section IV we show the experimental data obtained through the use of Strong Modulating Pulses (SMP) \cite{fortunato} and QST for creating computational base states, logic gates and quantum algorithms.

\section{Density matrix tomography method}

In this section we will revise the QST method in details, with special emphasis on the general features that makes the method suitable for use in systems with arbitrary spins.
 
\subsection{Magnetization and the deviation density matrix}

In a NMR experiment, the observables are the components of the macroscopic nuclear magnetization, which are proportional to the average values of the nuclear spin operators $\left< I_{\alpha}\right>$, $\alpha = x,y,z$:

\begin{equation}
M_{\alpha}=\left< \mu_{\alpha}\right> \propto \mbox{tr}(\rho\,I_{\alpha})
\end{equation}
where $\rho$ is the density operator. In the high temperature approxiamtion, $\hbar \omega_{L} \ll k_{B}T$, $\rho$ can be written as:
\begin{equation}\label{rho}
\rho \approx \frac{1}{2^n}\mathbb{I} + \epsilon\Delta\rho
\end{equation}
$n$ being the number of qubits, $\epsilon=\hbar\omega_{L}/k_{B}T \sim \ {10^{-5}}$ is the ratio between the magnetic and thermal energies of the system, and $\Delta\rho$ the traceless deviation matrix. In a quadrupolar spin system, the time evolution of the system magnetization, $M\left(t\right)$, is obtained considering the following propagator
\begin{equation}
U=e^{-\frac{i\mathcal{H}t}{\hbar}}
\end{equation}
in which
\begin{equation}
\mathcal{H}=\mathcal{H}_{Z}+\mathcal{H}_{Q},
\end{equation}
$\mathcal{H}_{Z}$ and $\mathcal{H}_{Q}$ are the Zeeman and the quadrupolar hamiltonian respectively. This hamiltonian is diagonal on the $\mbox{\bf I}^{2}$
and $I_{z}$ eigenstates basis. Thus, because only the deviation matrix in Eq.(\ref{rho}) evolves under the influence of the propagator $U$,
\begin{equation}
\Delta\rho\left(t\right)=U\,\Delta\tilde{\rho}\,U^{\dagger}.
\end{equation}
Where we considered $\Delta\rho(0)=\Delta\tilde{\rho}$ (see Appendix \ref{appa} for details).

The expression for the evolution of the transverse magnetization can be obtained from the expected
value of the $I_{+}$ operator. However, for a reason that will be
clarified later, an $\alpha$  phase factor will be considered:
\begin{equation}
M\left(t\right)=\mbox{tr}\left\{ U\,\Delta\tilde{\rho}\,U^{\dagger}I_{+}\right\} e^{i\alpha}\label{magt1}
\end{equation}
In Appendix \ref{appa} we show that $\Delta\tilde{\rho}$ can be expanded in the polarization operator $T_{l,m}$ basis as:
\begin{equation}
\Delta\tilde{\rho} =  \sum_{l,m,m'}a_{l,m}e^{i\left(\phi-\frac{\pi}{2}\right)\left(m-m'\right)}d_{m',m}^{l}\left(-\theta\right)T_{l,m'}\label{rhorot}.
\end{equation}
Thus, inserting Eq.(\ref{rhorot}) into Eq.(\ref{magt1}) and using the trace properties, we have:
\begin{equation}
M\left(t\right)= \sum_{l,m,m'} a_{l,m}e^{i\left(\phi-\frac{\pi}{2}\right)\left(m-m'\right)}d_{m',m}^{l}\left(-\theta\right)\mbox{tr}\left\{T_{l,m'}\,U^{\dagger}\,I_{+}\,U\right\} e^{i\alpha}.
\end{equation}
This expression can be simplified using the properties of the polarization operators $T_{l,m}$ \cite{Varsh}(see Appendix \ref{appb}), 
\begin{equation}
\left[T_{l,m'}\right]_{r,s}=\left[T_{l,m'}\right]_{r,s}\delta_{r,s+m'}.
\end{equation}
Using that
\begin{equation}
\left[I_{+}\right]_{s,r}=\left[I_{+}\right]_{s,r}\delta_{s,r+1},
\end{equation}
we obtain, for the delta product
\begin{equation}
\delta_{s,r+1}\delta_{r,s+m'} =\delta_{m',-1},
\end{equation}

that simplifies the magnetization to

\begin{equation}
M\left(t\right)=\sum_{l,m} a_{l,m} e^{i\left(\phi-\frac{\pi}{2}\right)\left(m+1\right)+i\alpha}d_{-1,m}^{l}\left(-\theta\right)\sum_{r,s}\lambda_{s}^{*}\lambda_{r}\left[I_{+}\right]_{s,r}\left[T_{l,-1}\right]_{r,s}.
\end{equation}
Where $\lambda_{s}= e^{-i\frac{E_{s}}{\hbar}t}$, and the $\lambda_{s}^{*}\lambda_{r}=e^{-i\frac{E_{r}-E_{s}}{\hbar}t}$
provides the oscillation with Bohr frequencies.

The term $d_{-1,m}^{l}\left(-\theta\right)$ 
is the reduced Wigner function, which gives the dependence on the nutation angle $\theta$. Hence, it is possible to maximize the magnetization by choosing $\theta$ that make $d_{-1,m}^{l}$ maximum. Using
the properties (\ref{propwig}), (\ref{proptlm}), and the hermiticiy of the density operator, we obtain

\begin{equation}
\ a_{l,m} = (-1)^m a_{l,-m}^*,
\end{equation}
it is possible to make the index change $m\rightarrow-m$ (the summation of $m$ runs over a symmetrical range, $\left|m\right|\leq l$), so the final magnetization expression is reached:

\begin{equation}
M\left(t\right)=\sum_{l,m}a_{l,m}^{*}e^{i\left(1-m\right)\left(\phi-\frac{\pi}{2}\right)+i\alpha}d_{1,m}^{l}\left(-\theta\right)\sum_{r,s}\lambda_{s}^{*}\lambda_{r}\left[I_{+}\right]_{s,r}\left[T_{l,1}^{\dagger}\right]_{r,s}.\label{magnet}
\end{equation}

From the definitions:

\begin{equation}
f_{s,r}\equiv\lambda_{s}^{*}\lambda_{r},
\end{equation}

\begin{equation}
\left[A_{l}\right]_{s,r}\equiv\left[{I}_{+}\right]_{s,r}\left[{T}_{l,1}^{\dagger}\right]_{r,s},
\end{equation}

and

\begin{equation}
S_{s,r}\equiv\sum_{l,m}a_{l,m}^{*}e^{i\left(1-m\right)\left(\phi-\frac{\pi}{2}\right)+i\alpha}d_{1,m}^{l}\left(-\theta\right)\left[A_{l}\right]_{s,r},\label{Ssrdef}
\end{equation}
we may write

\begin{equation}
M\left(t\right)=\sum_{s,r}f_{s,r}S_{s,r}.\label{Mtsimples}
\end{equation}


\subsection{Density matrix tomography}

With the magnetization written in terms of a sum involving the contributions of different orders of density matrix coherences, it is possible to design a coherence selection scheme, i.e. to write the NMR signal in terms of a single coherence order. Thus, the coherence selection was
done with the temporal average of several signals with the form

\begin{equation}
\overline{S}_{r,s}=\frac{1}{N_{p}}\sum_{n=0}^{N_{p}-1}S_{r,s}\left(\phi_{n},\alpha_{n}\right).\label{media1}
\end{equation}
Inserting Eq.(\ref{Ssrdef}) in Eq.(\ref{media1}), we obtain

\begin{equation}
\overline{S}_{r,s}=\frac{1}{N_{p}}\sum_{n=0}^{N_{p}-1}\sum_{l,m}a_{l,m}^{*}e^{i\left(1-m\right)\left(\phi_{n}-\frac{\pi}{2}\right)+i\alpha_{n}}d_{1,m}^{l}\left(-\theta\right)\left[A_{l}\right]_{s,r}.\label{Srsmedio}
\end{equation}

With the angles $\phi_{n}$ and $\alpha_{n}$ parametrized as

\begin{eqnarray}
\phi_{n} & = & 2\pi\frac{n}{N_{p}}+\frac{\pi}{2}, \nonumber\\
\alpha_{n} & = & 2\pi n\frac{\left(m'-1\right)}{N_{p}}\label{eq:},
\end{eqnarray}
the summation over $n$ simplifies to

\begin{equation}
\sum_{n}e^{2\pi i \frac{n}{N_{p}}\left(m'-m\right)}=N_{p}\delta_{m,m'},\; N_{p}\geq1+m'
\end{equation}
and Eq. (\ref{Srsmedio}) reduces to

\begin{equation}
\overline{S}_{r,r+1}\left(m'\right)=\sum_{l}a_{l,m'}^{*}d_{1,m'}^{l}\left(-\theta\right)\left[A_{l}\right]_{r,r+1}.
\end{equation}

It becomes evident now why the phase $\alpha$  was inserted in Eq. (\ref{magt1})
to permit the emergence of the Kronecker delta. Notice that has to be set to the correct value during the signal. Since the operator
$T_{0,0}\propto\mathbb{I}$ is not accessible through NMR experiments, the latter equation will be rewritten with the auxiliar variable $l'$:

\begin{eqnarray}
\overline{S}_{r,r+1}\left(m'\right) & = & \sum_{l=l'}^{2S} a_{l,m'}^{*}d_{1,m'}^{l}\left(-\theta\right)\left[A_{l}\right]_{r,r+1}\nonumber \\
l' & = & \{1,m'\}.\label{Srr1}
\end{eqnarray}
This expression corresponds to the linear system

\begin{eqnarray}
AX & = & B\label{AXB}\end{eqnarray}
with

\begin{eqnarray}
\left[A\right]_{i,l-l'+1} & = & \left[A_{l}\right]_{i,i+1},\nonumber \\
\left[X\right]_{l-l'+1} & = & a_{l,m'}^{*}d_{1,m'}^{l}\left(-\theta\right),\label{sistema}\\
\left[B\right]_{i} & = & \overline{S}_{i,i+1}\left(m'\right),\nonumber 
\end{eqnarray}
where

\begin{eqnarray}
i & = & 1,2,...,2S\nonumber \\
l & = & l',l'+1,...,2S.\label{il}
\end{eqnarray}

Therefore, for each average value, the respective coefficients
$a_{l,m'}$ (with $m'$ fixed) can be found. Because of the density operator hermiticity, only the coefficients for $m'\geq0$ need to be considered. To solve the system of equations, the least mean square method was used .

\section{Experimental tomography tests}

In order to verify the efficiency of the tomography method for quadrupolar nuclei and the influence of experimental
aspects in the method performance, some simple tests were performed. Those tests refer to the quadrupolar interaction influence
of the sample, inhomogeneity of the radio-frequency field, and possible imprecision arising from the pulse calibration process.
The sample used was cesium pentadecafluorooctanoate (CsPFO) and the nuclei observed was Cesium-133, wich has a spin of 7/2. The sample was dissolved in heavy water to the concentration of 37,5\%. When the sample is placed in the magnet the liquid crystal miscels align with the field and, then, display an aerage internal electric field gradient. This will result in the following quadrupolar hamiltonian for the Cesium nuclei:
\begin{equation}
H_{q}=\frac{\hbar\omega_{q}}{6}\left(3I^{2}_{z}-I^{2}\right).
\end{equation}

The $\omega_{q}$ is the quadrupolar frequency and its value is highly dependent on the sample temperature. Therefore, the sample temperature must be kept constant. The measurements were performed on a 400 MHz magnet using a 5 mm CP/MAS probe and a VARIAN Inova Unity spectrometer.

\subsection{Equilibrium, ${I}_{y}$ and even order coherence states creation}

In the base of ${T}_{l,m}$, the index $m$ refers to contributions over the density operator for the $m$ super-diagonal elements. For example, the $m=1$ term has only the first superdiagonal terms non-null, independently of the rank $l$. 
The first test described is the acquisition of the equilibrium
state ${I}_{z}$, which has all off diagonal null elements (it is proportional to ${T}_{1,0}$). In order to do so, we only need to keep the sample in the static magnetic field long enough for the spins to allign with it and then apply the tomography pulses. In the next test, the ${I}_{y}$ state is created by applying a $(\frac{\pi}{2})_{-x}$ pulse on the equilibrium state. As ${I}_{y}$ have only the first (imaginary) sub and super-diagonal non-null elements, after the tomography sequence, only the spectrum corresponding to the first coherence ($m=1$) will have non-zero intensities. The even order coherence test consists of creating states which have only pre-determined coherence orders by applying a pulse sequence composed of hard pulses and free evolution intervals. First, it is necessary to demonstrate that the free evolution changes the rank $l$, while the pulses change
the coherence order $m$. To go from $m$ to $m'$,
it is required that first the rank be changed from $l$ to $m'$, always
obeying the constraints for the ${T}_{l,m}$:

\begin{eqnarray*}
m & = & -l,-l+1,...,l-1,l\\
l & = & 0,1,...,2S.\end{eqnarray*}

\subsubsection{Changing the rank}

For free evolutions, in the rotating frame on resonance, the evolution operator will be

\begin{equation}
{U}_{e}=e^{-i\frac{\mathcal{H}_{Q}}{\hbar}t}\,,\,\mathcal{H}_{Q}=\frac{\hbar\omega_{Q}}{6}\left[3I_{z}^{2}-I(I+1)\mathbb{I}\right].
\end{equation}

Here, if the ${T}_{l,m}$ operator evolves under
the effect of ${U}_{e}$, the resulting $\widetilde{T}_{l,m}$
will be given by

\begin{equation}
\tilde{T}_{l,m}={U}_{e}{T}_{l,m}{U}_{e}^{\dagger}=\sum_{l'}b_{l,l'}{T}_{l',m}.
\end{equation}

Thus, the information about the presence of the ${T}_{l',m}$
component is in the $b_{l,l'}$ coefficient, wich may be evaluated by the projection of $\tilde{T}_{l,m}$ over ${T}_{L,m}$:

\begin{equation}
b_{l,L}=tr\left\{ {T}_{L,m}^{\dagger}e^{-i\frac{\mathcal{H}_{Q}}{\hbar}t}{T}_{l,m}e^{i\frac{\mathcal{H}_{Q}}{\hbar}t}\right\}
\end{equation}

For the rank transfer experiments, it is necessary to find the free evolution time ${t}_{e}$ for which the absolute value of $b_{l,L}$ is maximum. For producing a state with only even order coherences this time is ${t}_{e}=\frac{\pi}{{\omega}_{Q}}$. 

\subsubsection{Changing the order}

Using $(-\frac{\pi}{2})_{-y}$ pulses, based on the properties of rotations, and the fact that the density
operator has contributions in the ${T}_{l,m}+{T}_{l,m}^{\dagger}$
format, we have (see Eq. (\ref{rottlm})):

\begin{eqnarray*}
{D}\left({T}_{l,m}+{T}_{l,m}^{\dagger}\right){D}^{\dagger} & = & \left[\left(-1\right)^{m}d_{0,m}^{l}\left(-\theta\right)+d_{0,-m}^{l}\left(-\theta\right)\right]{T}_{l,0}+\\
 & + & \sum^{l}_{m'=1}\left[\left(-1\right)^{m'}d_{m',-m}^{l}\left(-\theta\right)+\left(-1\right)^{m-m'}d_{m',m}^{l}\left(-\theta\right)\right]\left({T}_{l,m'}+{T}_{l,m'}^{\dagger}\right),
\end{eqnarray*}

\begin{figure}
	\centering
		\includegraphics[width=0.70\textwidth]{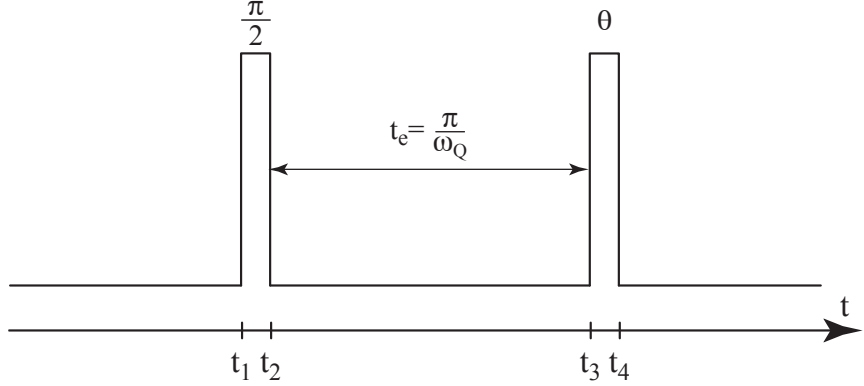}
	\caption{Pulse sequence for the even order coherence creation}
	\label{fig:fig1}
\end{figure}

\begin{figure}
	\centering
		\includegraphics[width=0.70\textwidth]{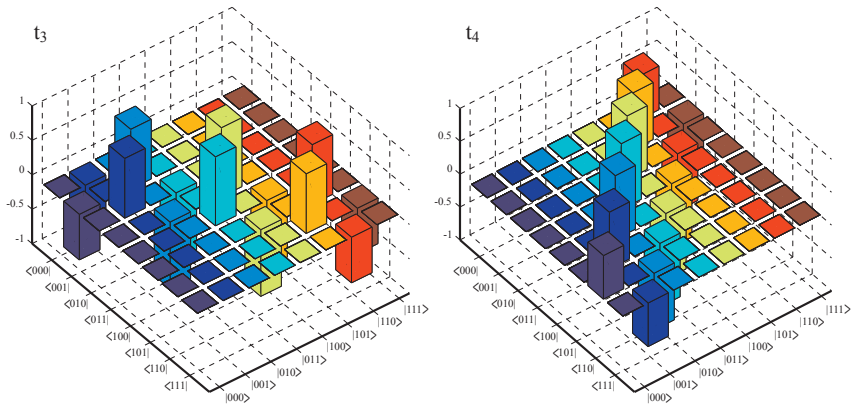}
	\caption{Real parts of intermediate ${t}_{3}$ and final ${t}_{4}$ states of the pulse sequence shown on figure \ref{fig:fig1}. All intensities normalized.}
	\label{fig:fig2}
\end{figure}

As showed by this expression, under unitary operations (i.e., rotations), the polarization operators ${T}_{l,m}$ remains with the rank $l$ unchanged. For the even order coherence state creation, it is necessary to find the $theta$ angle value on the expression above for what the contributions by odd $m$ are minimized (or null) and the even $m$ maximized. By analysing the several contributions to the summation for $m'=1...l$, it was found that the even order maximizations occurs when $\theta\ = \frac{\pi}{2} $. 
In figure \ref{fig:fig1} the pulse sequence for the even order coherence creation is shown. The intermediate and final states are shown on figure \ref{fig:fig2}.

\subsection{Tomography Results}

As discussed before, the reduced Wigner functions $d_{1,m}^{l}\left(-\theta\right)$
are related to the nutation angle of each component and, in
the tomography process, it is convenient that their values are as high as possible. 
Thus, the angle for which those functions are maximized
should be the more convenient choice. However, even with the change to Eq. (\ref{magnet}), that can lead to long pulses,
making the pulse imperfections more severe and the tomography process less precise because selective excitation effects can start to show in the pulses mainly for large quadrupolar coupling. 
An alternative is to find angles that not necessarily maximize the
$d_{1,m}^{l}\left(-\theta\right)$ function in particular, but those that
for a given $m$ make the values of these functions in relation to two or more values of $l$ high enough for a considerable sensibility for this component. Thus, we used the tomography angles on the table \ref{tab1}, which were selected based on this criteria.
In the test analysis the following convention was adopted: $FE\equiv$ fidelity between the experimental state and its theoretical equivalent; $FS\equiv$ fidelity between the experimental state and the simulation.
For the simulations, a $\varepsilon$ parameter related with the pulse precision was multiplied by the pulse intensity. A gaussian shape was adopted for the RF magnetic oscilating field, giving it a $5\%$ maximum variation in intensity.

\begin{table}
\caption{Nutation angle in radians for the tomography pulses} \label{tab1} \centering
\begin{tabular}{c|c|c|c|c|c|c|c|c|}

\cline{1-9}
\multicolumn{1}{|c|} l & m=0 & m=1 & m=2 & m=3 & m=4 & m=5 & m=6 & m=7 \\
\cline{1-9}
\multicolumn{1}{|c|} 1 & {\multirow{2}{*}{0.960}} & {\multirow{7}{*}{0}} & - & - & - & - & - & - \\
\cline{1-1}
\cline{4-9}
\multicolumn{1}{|c|} 2 & {} & {} & {\multirow{3}{*}{0.606}} & - & - & - & - & - \\
\cline{1-2}
\cline{5-9}
\multicolumn{1}{|c|} 3 & {\multirow{2}{*}{0.462}} & {} & {} & 1.230 & - & - & - & - \\
\cline{1-1}
\cline{5-9}
\multicolumn{1}{|c|} 4 & {} & {} & {} & {\multirow{2}{*}{0.680}} & {\multirow{2}{*}{1.020}} & - & - & - \\
\cline{1-2}
\cline{4-4}
\cline{7-9}
\multicolumn{1}{|c|} 5 & {\multirow{3}{*}{0.268}} & {} & {\multirow{3}{*}{0.292}} & {} & {} & {\multirow{2}{*}{1.094}} & - & - \\
\cline{1-1}
\cline{5-6}
\cline{8-9}
\multicolumn{1}{|c|} 6 & {} & {} & {} & {\multirow{2}{*}{0.426}} & {\multirow{2}{*}{0.604}} & {} & 1.404 & - \\
\cline{1-1}
\cline{7-9}
\multicolumn{1}{|c|} 7 & {} & {} & {} & {} & {} & 0.730 & 0.928 & 1.426 \\
\cline{1-9}
\end{tabular}
\end{table}

The state with coherences of even orders was chosen as the guide for the simulations due to the fact that its expansions have components over a greater number of ${T}_{l,m}$ operators. With the $FE$
determined, $\varepsilon$ and the quadrupolar frequency values were inserted on the
simulation to reach a state with $FS\approx FE$. Therefore, the choice
of these parameters was based on the experimental conditions, given a value of $\varepsilon = 0.95$.
When the parameters were defined, they were also used for the
${I}_{z}$ and ${I}_{y}$ simulations. 

\begin{table}
\caption{Experimental fidelities (in \%) for each test.} \label{effet} \centering
\begin{tabular}{|c|c|c|c|}
\hline
{} & $I_{z}$ & $I_{y}$ & Even order coherences\\
\hline
FE & 99,66 & 81,50 & 75,25 \\
\hline
FS & 99,28 & 97,73 & 76,12 \\
\hline
\end{tabular}
\end{table}

\begin{figure}
	\centering
		\includegraphics[width=0.70\textwidth]{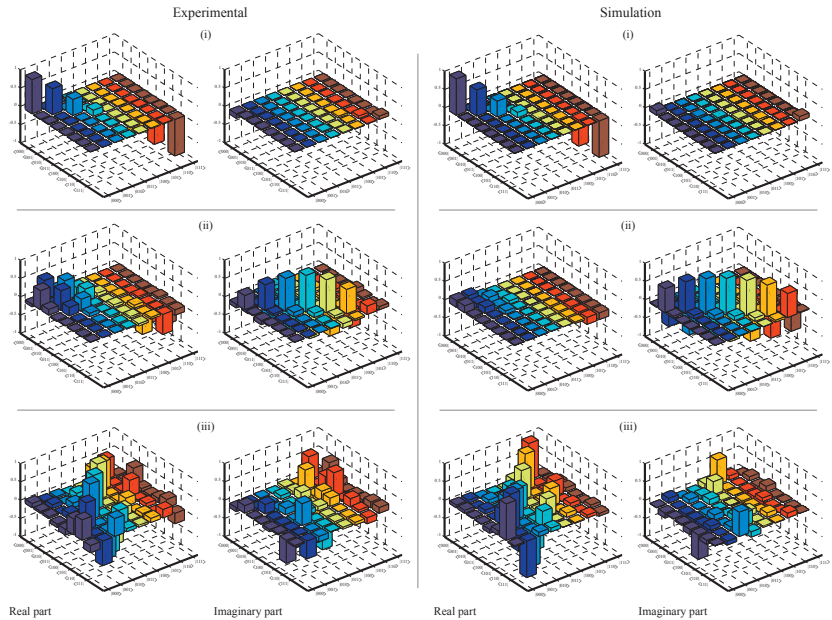}
	\caption{At right, experimental data and, on left, simulation of the equilibrium state ${I}_{z}$(i), ${I}_{y}$(ii) and even coherence order state (iii). All intensities are normalized.}
	\label{fig:fig3}
\end{figure}

The experimental results and the simulations are shown on the \ref{fig:fig3} and the fidelities are displayed on table \ref{effet}.  

To acquire well-defined spectra,
we chose for each coherence an average number obeying the progression
$128,144,...,240$. This gives the tomography procedure a duration on the
order of 1 hour and 30 minutes. By a visual analysis of the experimental
results and their respective simulations, it is possible to see that
the matrix shape was well reproduced.

\section{SMP Results}

\subsection{Computational base states using SMP}

\begin{figure}[h!]
\centering
\includegraphics[height=15cm]{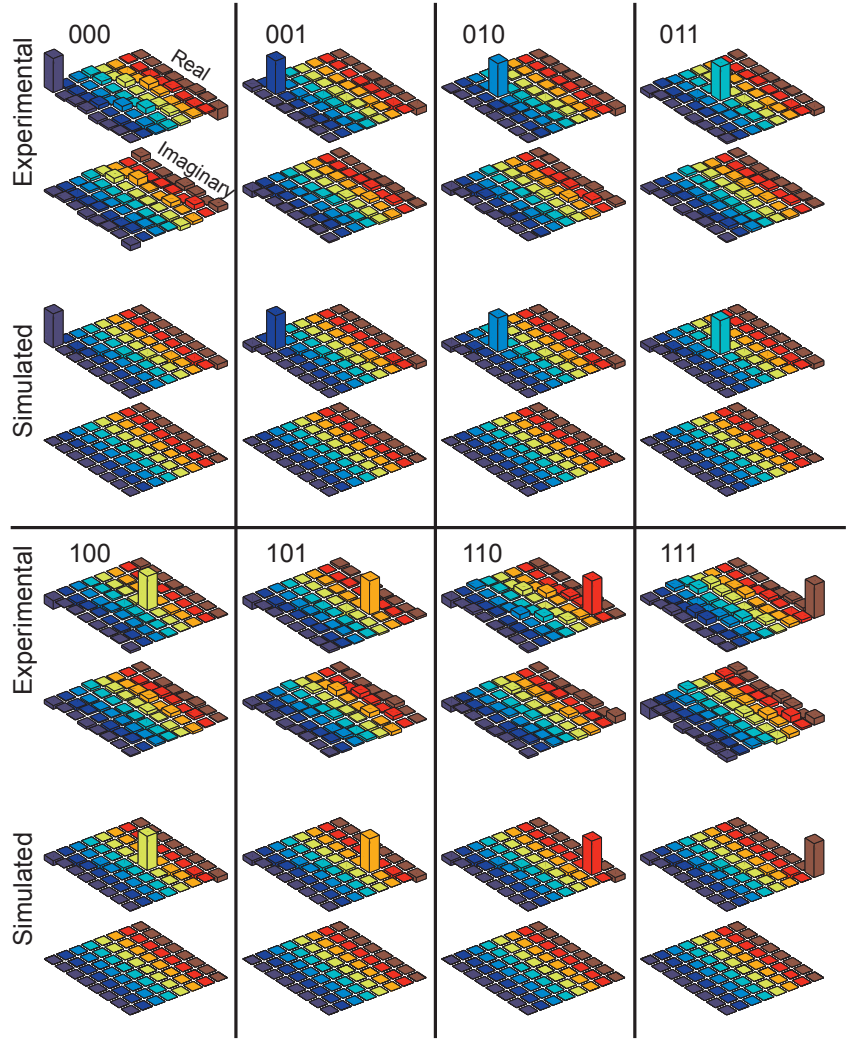}
\caption{\label{figpps} Experimental density matrices and their respective expected simulated density matrices}
\end{figure}

Having confirmed the effectiveness of the tomography method, we now have a tool to determine completely the quantum state of the cesium nuclei in our sample. We wish then to demonstrate the possibility of performing a complete quantum algorithm in this system using the SMP technique \cite{fortunato}. However, before the implementation of the quantum algorithm we wish to demonstrate the construction of the pseudo-pure states 
used as the computational base. They will be the starting point of every algorithm.
First we optimized the SMP pulse sequences for the preparation of the computational base states from $\left|000\right\rangle$ to $\left|111\right\rangle$. The optmization was designed to employ four temporal averages, each one with ten segments. In Fig. \ref{figpps}, the experimental density matrices obtained through the quantum state tomography process are shown for each state, also with the corresponding simulated matrices that were expected.

\begin{table}
\caption{Fidelity of the computational base states created} \label{fidpps}
\centering
\begin{tabular}{c|c}
\hline State & Experimental fidelity (\%) \\ \hline \hline
$\left|000\right\rangle$ & 82,5 \\
\hline
$\left|001\right\rangle$ & 93,8 \\
\hline
$\left|010\right\rangle$ & 95,2 \\
\hline
$\left|011\right\rangle$ & 95,8 \\
\hline
$\left|100\right\rangle$ & 92,8 \\
\hline
$\left|101\right\rangle$ & 88,8 \\
\hline
$\left|110\right\rangle$ & 91,1 \\
\hline
$\left|111\right\rangle$ & 87,2 \\
\hline
\end{tabular}
\end{table}

Table \ref{fidpps} gives the experimental fidelities corresponding to the density matrices presented in Fig. \ref{figpps}. The fidelities found during the optmization of the SMPs are very close to unity, but
as it can be seen, the experimental fidelities will always have a smaller value. That can be caused by experimental errors such as the ones described in previous section: a) pulse calibration; b) lack of spatial homogeneity of the $B_{1}$ field of the RF pulses; c) limitations of the 
temporal, phase and amplitude resolution of the spectrometer; d) errors due to the response time of the RF coil during the rapid variations of power levels required by the SMP; e) relaxation processes during the pulse application. 

Among these different types of erros, there are some that are more influencing on the final result. From the simulations it was possible to determine which ones were more effective in reducing the fidelity of the measurements.
The calibration errors are a strong factor in the results. As demonstrated by the simulations, a $5\%$ calibration error may result in a $30\%$ loss in fidelity. This shows the importance of a very precise pulse calibration. The inhomogeneity of the RF pulses is another important factor, but in this case not as much as 
the RF coil response time. The rapid variation of the SMPs amplitudes (sharp edges) is not sometimes well reproduced by the RF coil and this greatly reduces the fidelity. One way around this is forcing an SMP optimization that 
favors sequences with smooth amplitude transitions.

An important thing to notice about the use of the SMP here is that the pulse durations for the creation of these states lie between 100 to 300 $\mu$s. If transition-selective pulses were used, the duration would be of the order of miliseconds.

\subsection{Logic Gates Implementation using SMP}

The next set of experiments we present consist of the logic gates implementation. For that purpose, we created a SMP that represents a logic gate and applied it to some of the computational base states. Then, we used the quantum state
tomography method to check the output of the logic gate.
The SMP pulse sequences for the gates were optimized with 25 segments and the optmization process took about 10 minutes to reach fidelities over 99.99\%. We chose to present here the results obtained for the Toffolli gates, since they are
more graphically representative.

\begin{figure}[h!]
\centering
\includegraphics[height=8.5cm]{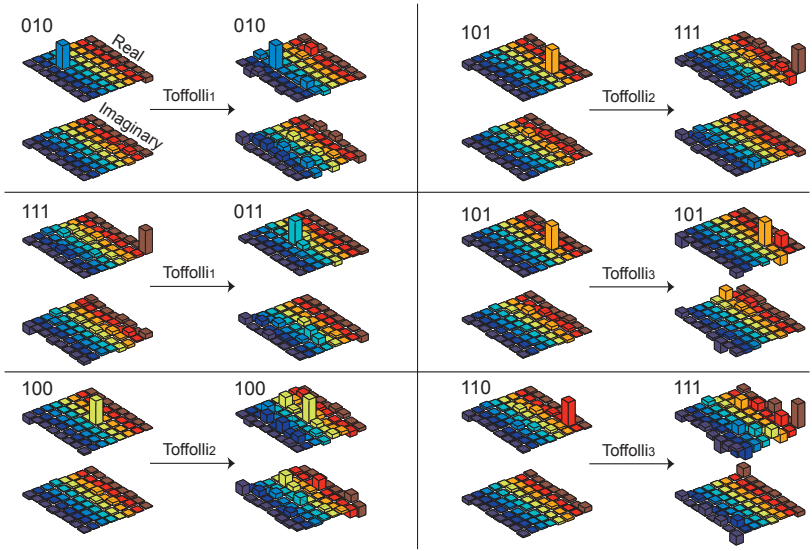}
\caption{\label{111tof1} Experimental density matrices of some computational base states before and after the application of the Toffolli gates with different target q-bits.}
\end{figure}

In Fig. \ref{111tof1} we show the tomographed density matrices before and after the application of the different Toffolli gates on the states $\left|000\right\rangle$,
$\left|001\right\rangle$, $\left|010\right\rangle$,
$\left|100\right\rangle$, $\left|011\right\rangle$ and $\left|111\right\rangle$. The experimental fidelities are shown in Table \ref{fidportas}.

\begin{table}
\caption{Toffolli gate fidelities.} \label{fidportas} \centering
\begin{tabular}{c|c}
\hline Operation & Fidelity (\%) \\ \hline \hline
Toffolli$_{1}\left|010\right\rangle$ & 63,3 \\
\hline
Toffolli$_{1}\left|111\right\rangle$ & 69,7 \\
\hline
Toffolli$_{2}\left|100\right\rangle$ & 62,6 \\
\hline
Toffolli$_{2}\left|101\right\rangle$ & 66,2 \\
\hline
Toffolli$_{3}\left|110\right\rangle$ & 43,3 \\
\hline
Toffolli$_{3}\left|101\right\rangle$ & 60,3 \\
\hline
\end{tabular}
\end{table}

As can be seen, despite the low fidelities, the final state can still be easily recognized as the one expected after the gate application. This demonstrates the correct performance of the SMP.
To perform logic gates using transition-selective pulses on a $I=7/2$ nuclear spin system, about 2 ms are required for the pulse duration. The SMP we used were at most 300 $\mu$s long and therefore the 10 ms relaxation time for our sample has a negligible effect.

\subsection{Deutsch-Jozsa Algorithm}

\begin{figure}[h!]
\centering
\includegraphics[height=6cm]{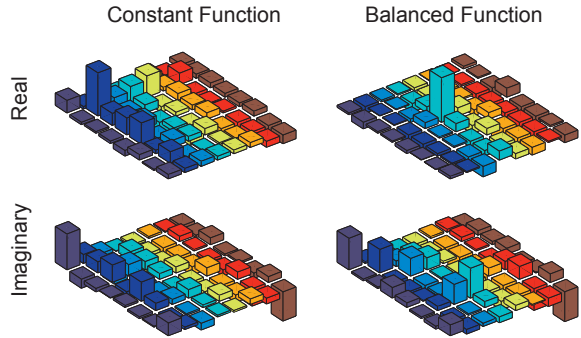}
\caption{\label{djexp} Experimental density matrices obtained after the application of the Deutsch-Jozsa algorithm for a case of a constant and a case of balanced function. The final states should be $\left|001\right\rangle$ (constant) and
$\left|011\right\rangle$ (balanced)}
\end{figure}

With the use of SMPs it was possible to implement the Deutsch-Jozsa algorithm \cite{deutsch}. Here we present the result for two cases: one in which the tested function is balanced and one in which it is constant. In the case of a 
constant function the final result will be the state $\left|000\right\rangle$ or $\left|001 \right\rangle$ and any other state if the function is balanced. In Fig. \ref{djexp}  the two results are shown. The constant function corresponded
to one that always returned 1 for any input state. The balanced function returned 0 for the states $\left|00\right\rangle$ and $\left|10\right\rangle$ and returned 1 for the states $\left|01\right\rangle$ e $\left|11\right\rangle$.
Thus, the final state of those constant and balanced functions is $\left|001\right\rangle$ and $\left|011\right\rangle$ respectively. We can visually recognize that the expected states are found in the experimental data, demonstrating the successful
application of the SMP.

Table \ref{fiddj} displays the experimental fidelity of the data obtained. The low fidelity values are attributed to the many experimental erros involved. Still, the final state is easily indentified. If conventional pulses were used, a sequence of at least 5 ms would be necessary to perform the algorithm, which would render it useless due to relaxation effects. Also a great analytical effort would be required
to design a sequence that was the most efficient as possible.

\begin{table}
\caption{Fidelity of the results obtained after the Deutsch-Jozsa SMP application} \label{fiddj}
\centering
\begin{tabular}{c|c}
\hline State & Experimental Fidelity (\%)\\ \hline \hline
Deutsch-Jozsa $f_{c}$ & 64,3 \\
\hline
Deutsch-Jozsa $f_{b}$ & 65 \\
\hline
\end{tabular}
\end{table}

\subsection{Grover Algorithm}

\begin{figure}[h!]
\centering
\includegraphics[height=6cm]{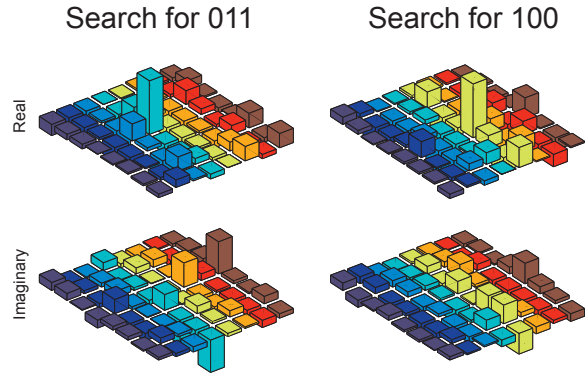}
\caption{\label{groverexp} Experimental density matrices after the application of the Grover algorithm in search of the states $\left|011\right\rangle$ and $\left|100\right\rangle$}
\end{figure}

Our last result was the implementation of the Grover algorithm \cite{groveralg}. This is an algorithm with practical uses, in which there is a $\sqrt{N}$ order gain compared to the classical equivalent. Figure \ref{groverexp} shows the experimental data
obtained for two cases of the algorithm. In the first case the search was for the $\left|011\right\rangle$ state and in the second case the search was for the $\left|100\right\rangle$ state.

Theoretically, the final state of this algorithm for 3 qubits is not a computational base state. In theory, the final result has a $97\%$ probability that it will return the state we were searching for. Amidst all our experimental errors,
this is of little concern in our final results. On the experimental data, we can identify the states we were looking for even though the fidelities were not as high as desired. The fidelities are recorded in Table \ref{fidgrover}. 
Due to all the experimental errors involved, these fidelities are as high as expected.

\begin{table}
\caption{Fidelities for the Grover algorithm} \label{fidgrover}
\centering
\begin{tabular}{c|c}
\hline State & Experimental (\%)\\ \hline \hline
Grover$_{011}$ & 52 \\
\hline
Grover$_{100}$ & 52 \\
\hline
\end{tabular}
\end{table}

We were able to check the performance of the SMP once again for another algorithm. Just as in the earlier cases, if we were to employ conventional pulses we would need a very long pulse sequence (greater than the relaxation time)
and also a great analytical effort to design it. 

\section{Conclusions}

The QST method applied met the requirements of duration and precision of our system. That could be shown from the small differences between the simulated and experimental results. This could be evidenced by the tomography of the following states:

\begin{itemize}
\item ${I}_{z}$: The result is close to perfection, but
this state has only coherences of 0 order, and therefore it is not a good
parameter for the tomography method analysis.
\item $\hat{I}_{y}$: Experimentally, it is possible to see the emerging
of contributions on the real part with coherences of order \emph{1}
and, on the imaginary part, of coherences of order 2 (with much smaller
intensities); yet, the most intense elements are exactly those
that we expect theoretically.
\item \emph{Even order coherence:} The elements that were theoretically expected can all be seen. The simulations
with the introduction of the experimental imperfections, reproduced
very well the real part observed experimentally.
\end{itemize}

The construction of the computational base states was achieved with fidelity greater than $80\%$, which is a good indication of the efficiency of the method. In the case of logic gates, the fidelities fall to the order of $60\%$, as expected due to the
concatenation of SMP. In the same way, we find fidelities between $50\%$ and $60\%$ for the quantum algorithms.
We have successfully shown the advantages and experimental applications of SMP for quantum information processing in 3 q-bit systems composed of 7/2 spin nuclei. The experimental data shows us that SMP allow us to perform logic operations
in a time scale ten times smaller than with conventional pulses \cite{bonkjmr}. The implementation of complete quantum algorithms was only made possible through the use of SMP. All the data was acquired through the quantum state tomography
sequence which also had its efficiency tested. Both SMP and the quantum state tomography have many applications in NMR outside the scope of quantum information processing. SMP can make designing a pulse sequence a much simpler task
without any analitical effort and sequences with much shorter duration and selectivity can be achieved through the use of SMP.

\begin{acknowledgments}
We would like to thank FAPESP, CAPES and CNPQ for financial support throughout this reasearch.
\end{acknowledgments}

\newpage

\appendix
\setcounter{equation}{0}


\section{}\label{appa}

\subsection{Rotations over the density operator}

To describe the rotations over the $\Delta\hat{\rho}$ operator, it
will be expand in an ortonormal basis formed by the polarization operators
$\left\{ {T}_{l,m}\left(S\right)\right\} $ \cite{Varsh}:

\begin{equation}
\Delta{\rho}=\sum_{l=1}^{2S}\sum_{m=-l}^{l} a_{l,m}{T}_{l,m}.
\end{equation}

Applying the rotation operator ${D}\left(\alpha,\beta,\gamma\right)$ to
the last expression, 

\begin{equation}
{D}\left(\alpha,\beta,\gamma\right)\Delta\hat{\rho}{D}^{\dagger}\left(\alpha,\beta,\gamma\right)=\sum_{l,m} a_{l,m}e^{-im\gamma}\sum_{m'}e^{-im'\alpha}d_{m',m}^{l}\left(\beta\right){T}_{l,m'}.
\end{equation}

\begin{figure}
	\centering
		\includegraphics[width=0.50\textwidth]{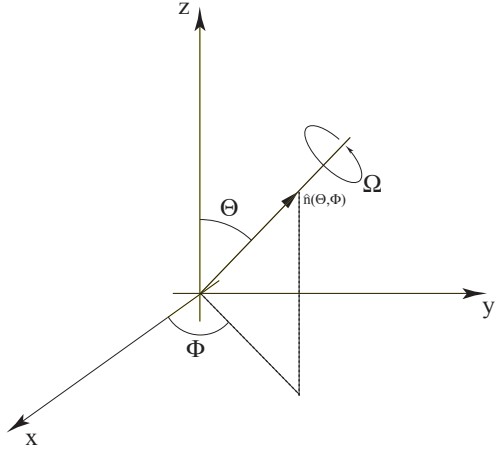}
	\caption{Simplified description of the rotations}
	\label{fig:rotacoes}
\end{figure}

Any rotation of a coordenate system which takes $\left(x,y,z\right)$
to $\left(x',y',z'\right)$ can be treated as a rotation of an 
angle $\Omega$ around the $n\left(\Theta,\Phi\right)$ axis, where $\Theta$
and $\Phi$ are the same in both systems \cite{Varsh}, as illustred in the figure \ref{fig:rotacoes}.

The Euler angles $\left(\alpha,\beta,\gamma\right)$ and the $\left(\Omega,\Theta,\Phi\right)$
angles are connected by:

\begin{eqnarray}
\cos\left(\frac{\Omega}{2}\right) & = & \cos\left(\frac{\beta}{2}\right)\cos\left(\frac{\alpha+\gamma}{2}\right),\nonumber \\
\tan\Theta & = & \frac{\tan\left(\frac{\beta}{2}\right)}{\sin\left(\frac{\alpha+\gamma}{2}\right)},\label{eq:}\\
\Phi & = & \frac{\pi}{2}+\frac{\alpha-\gamma}{2}.\nonumber
\end{eqnarray}

For a rotation axis constrained in the transversal plane, $\Theta=\frac{\pi}{2}$. If the nutation  axis makes a $\phi$ angle with the $x$ axis, $\Phi=\phi$. Then, for rotations of $\theta$ about the axis $\Omega=\theta$, we have:

\begin{eqnarray}
\gamma & = & -\alpha,\nonumber \\
\alpha & = & \phi-\frac{\pi}{2},\label{angulos}\\
\beta & = & -\theta\nonumber. 
\end{eqnarray}

In NMR experiments, $\theta$ indicates the rotation
associated to the pulse, $\Theta$ the angle between
the magnetic static field and the RF oscillating field (because they are perpendicular,
$\Theta=\frac{\pi}{2}$) and $\Phi$ gives the phase of
the RF field. Inserting Eq. (\ref{angulos}) in Eq. (\ref{Dlmm}):

\begin{eqnarray}
D_{m',m}^{l} & = & e^{i\left(\phi-\frac{\pi}{2}\right)\left(m-m'\right)}d_{m',m}^{l}\left(-\theta\right)\label{eq:}
\end{eqnarray}
Then, the expression of the operator$\Delta{\rho}$
turned into $\Delta\tilde{\rho}$, is obtained:

\begin{eqnarray}
\Delta\tilde{\rho} & = & \sum_{l}\sum_{m,m'}a_{l,m}e^{i\left(\phi-\frac{\pi}{2}\right)\left(m-m'\right)}d_{m',m}^{l}\left(-\theta\right){T}_{l,m'}\label{r0rot1}
\end{eqnarray}

\section{}\label{appb}
\subsection{Polarization operators ${T}_{l,m}$}

To tomograph the density matrix operator, it is necessary to expand the operator
in a convenient basis. The polarization operators ${T}_{l,m}\left(S\right)$
were chosen for basis due to their properties under rotations. On the next calculus, all the algebric properties were extracted from \cite{Varsh}. The ${T}_{l,m}$ operators
are defined by the Clebsch-Gordan $C_{l_{1},m_{1},l_{2},m_{2}}^{l,m}$
coefficients

\begin{equation}
\left[{T}_{l,m}\right]_{\sigma',\sigma}=\sqrt{\frac{2L+1}{2S+1}}C_{S,\sigma,l,m}^{S,\sigma'},
\end{equation}
where

\begin{eqnarray*}
\left|l_{1}-l_{2}\right|\leq & l & \leq l_{1}+l_{2},\\
m=m_{1}+m_{2} & = & -l,-l+1,...,l-1,l,\\
l & = & 0,1,...,2S,\\
\sigma,\sigma' & = & -S,-S+1,...,S,\\
\sigma'&=&\sigma + m ,\\
\end{eqnarray*}
and

\begin{equation}
\left[{T}_{l,-1}\right]_{r,s}=-\left[{T}_{l,1}^{\dagger}\right]_{r,s}.\label{proptlm}
\end{equation}

The $l$ index is the rank of the operator, where the $m$
index is the coherence order. The rotations ${D}\left(\alpha,\beta,\gamma\right)$
in the state space are descrited by the Wigner
functions, $D_{m,m'}^{l}$. When the rotation operates in an eingenstate $\left|l,m\right\rangle $ of both ${I}^{2}$ and ${I}_{z}$, it will result in a combination
of states with a common value of $l$:

\begin{equation}
{D}\left(\alpha,\beta,\gamma\right)\left|l,m\right\rangle =\sum_{m'=-l}^{l} D_{m',m}^{l}\left(\alpha,\beta,\gamma\right)\left|l,m'\right\rangle, 
\end{equation}
where $\alpha,\beta$ and $\gamma$ are the Euler angles and

\begin{eqnarray}
D_{m',m}^{l}\left(\alpha,\beta,\gamma\right) & = & e^{-i\left(m'\alpha+m\gamma\right)}d_{m',m}^{l}\left(\beta\right),\nonumber \\
d_{m',m}^{l}\left(\beta\right) & = & \left\langle l,m'\right|e^{-i\beta\hat{L}_{y}}\left|l,m\right\rangle \label{Dlmm}.
\end{eqnarray}
The $d_{m',m}^{l}\left(\beta\right)$ are the reduced Wigner
functions, which obeys:

\begin{equation}
d_{m,m'}^{l}\left(\beta\right)=\left(-1\right)^{m-m'}d_{-m,-m'}^{l}\left(\beta\right).\label{propwig}
\end{equation}

The Clebsh-Gordan coefficients $C_{l_{1},m_{1},l_{2},m_{2}}^{l,m}$
and the reduced Wigner functions $d_{m',m}^{l}\left(\beta\right)$
can be numerically calculated by the use of the following expressions:

\begin{eqnarray}
d_{m',m}^{l}\left(\beta\right) & = & \left[\left(l+m\right)!\left(l-m\right)!\left(l+m'\right)!\left(l-m'\right)!\right]^{\frac{1}{2}}\times\nonumber \\
 & \times & \sum_{k}\left(-1\right)^{k}\frac{\left[\cos\left(\frac{\beta}{2}\right)\right]^{2l-2k+m-m'}\left[\sin\left(\frac{\beta}{2}\right)\right]^{2k-m+m'}}{k!\left(l+m-k\right)!\left(l-m'+k\right)!\left(m'-m+k\right)!},\label{eq:}
\end{eqnarray}

\begin{eqnarray}
C_{a\alpha b\beta}^{c\gamma} & = & \frac{\delta_{\gamma,\alpha+\beta}}{\Delta\left(a,b,c\right)}\left[\frac{\left(a+\alpha\right)!\left(a-\alpha\right)!\left(c+\gamma\right)!\left(c-\gamma\right)!\left(2c+1\right)}{\left(b+\beta\right)!\left(b-\beta\right)!}\right]^{\frac{1}{2}}\times\nonumber \\
 & \times & \sum_{z}\frac{\left(-1\right)^{a-\alpha+z}\left(a+b-\gamma-z\right)!\left(b+c-\alpha-z\right)!}{z!\left(a-\alpha-z\right)!\left(c-\gamma-z\right)!\left(a+b+c+1-z\right)!},\label{eq:}
\end{eqnarray}
where

\begin{equation}
\Delta\left(a,b,c\right)=\left[\frac{\left(a+b-c\right)!\left(a-b+c\right)!\left(-a+b+c\right)!}{\left(a+b+c+1\right)!}\right].
\end{equation}

Therefore, over a polarization operator, the rotation acts like:

\begin{equation}
{D}\left(\alpha,\beta,\gamma\right){T}_{l,m}{D}^{\dagger}\left(\alpha,\beta,\gamma\right)=\sum_{m'=-l}^{l}e^{-i\left(m'\alpha+m\gamma\right)}d_{m',m}^{l}\left(\beta\right){T}_{l,m'}.\label{rottlm}
\end{equation}


\newpage

\begin{thebibliography}{2}

\bibitem{ref1} U. Leonhardt and H. Paul, Phys. Rev. A \textbf{52}, 4899 (1995).

\bibitem{ref2} U. Leonhardt, Phys. Rev. Lett \textbf{76}, 4293 (1996).

\bibitem{ref3} R. Walser, J. I. Cirac and P. Zoller, Phys. Rev. Lett \textbf{77}, 2658 (1996).

\bibitem{ref4} A. S. Parkins, P. Marte, P. Zoller, O. Carnal and H. J. Kimble, Phys. Rev. A  \textbf{51}, 1578 (1995).

\bibitem{ref5} J. P. Amiet and St. Weigert, J. Phys. A \textbf{32}, L269 (1999).

\bibitem{livro} I. Oliveira, T. J. Bonagamba, R. S. Sarthour, J. C. C. Freitas, E. R. deAzevedo, \emph{NMR quantum information processing} (Elsevier, Amsterdam, 2007).  

\bibitem{relax2} R. S. Sarthour, E. R. deAzevedo, F. A. Bonk, E. L. G. Vidoto, T. J. Bonagamba, A. P. Guimarães, J. C. C. Freitas and I. S. Oliveira, Phys. Rev. A \textbf{68}, 022311 (2003). 

\bibitem{CHUANG} I. L. Chuang, N. Gershenfeld, M. G. Kubinec and D. W. Leung, Proc. R. Soc. London, Ser. A \textbf{454}, 447 (1998).

\bibitem{Hong} F. Hong, J. Ye, L. Ma, S. Picard, C. J. Bordé and J. L. Hall, J. Opt. B: Quantum Semiclassical Opt. \textbf{3}, 376 (2001).

\bibitem{CHUANG2} I. L. Chuang, N. Gershenfeld and M. Kubinec, Phys. Rev. Lett. \textbf{80}, 3408 (1998).

\bibitem{daskumar} R. Das and A. Kumar, Phys. Rev. A \textbf{68}, 032304 (2003). 

\bibitem{KAMP} H. Kampermann and W. Veeman, Quantum Inf. Process. \textbf{1}, 327 (2002).

\bibitem{Khitrin} A. K. Khitrin and B. M. Fung, J. Chem. Phys. \textbf{112}, 6963 (2000). 

\bibitem{sinha} N. Sinha, T. S. Mahesh, K. V. Ramanathan and A. Kumar, J. Chem. Phys. \textbf{114}, 4415 (2001).

\bibitem{bonkjmr} F. A. Bonk, E. R. deAzevedo, R. S. Sarthour, J. D. Bulnes, J. C. Freitas, A. P. Guimarães, I. S. Oliveira and T. J. Bonagamba, J. Magn. Reson. \textbf{175}, 226 (2005).

\bibitem{bonkpra} F. A. Bonk, R. S. Sarthour, E. R. deAzevedo, J. D. Bulnes, G. L. Mantovani, J. C. C. Freitas, T. J. Bonagamba, A. P. Guimarães and I. S. Oliveira, Phys. Rev. A \textbf{69}, 0423221 (2004).

\bibitem{bulnes} J. D. Bulnes, F. A. Bonk, R. S. Sarthour, E. R. de Azevedo, J. C. C. Freitas, T. J. Bonagamba and I. S. Oliveira, Braz. J. Phys. \textbf{35}, 617 (2005). 

\bibitem{vega1} S. Vega, J. Chem. Phys. \textbf{68}, 5518 (1978).

\bibitem{vega2} S. Vega and A. Pines, J. Chem. Phys. \textbf{66}, 5624 (1977).

\bibitem{ernst} A. Wokaun and R. R. Ernst, J. Chem. Phys. \textbf{67}, 1752 (1977).

\bibitem{telesartigo} J. Teles, E. R. deAzevedo, R. Auccaise, R. S. Sarthour, I. S. Oliveira and T. J. Bonagamba, J. Chem. Phys. \textbf{126}, 154506 (2007).

\bibitem{SMPKAMP} H. Kampermann and W. S. Veeman, J. Chem. Phys. \textbf{122}, 214108 (2005)

\bibitem{relax1} A. M. Souza, A. G. Viana, I.S. Oliveira, R.S. Sarthour, R. Auccaise, E.R. deAzevedo and  T. J. Bonagamba, Quantum Inf. Comp. \textbf{10}, 653, (2010).

\bibitem{relax3} D. O. Soares-Pinto, L. C. Celeri, R. Auccaise, F. F. Fanchini, E. R. deAzevedo, J. Maziero, T. J. Bonagamba and R. M. Serra, Phys. Rev. A \textbf{81}, 062118 (2010). 

\bibitem{relax5} R. Auccaise, J. Teles, R. S. Sarthour, T. J. Bonagamba, I. S. Oliveira, E. R.  deAzevedo, J. Magn. Reson. \textbf{192}, 17 (2008). 

\bibitem{relax4} A. Gavini-Viana, A. M. Souza, D. O. Soares-Pinto, J. Teles, R. S. Sarthour, E. R. deAzevedo, T. J. Bonagamba and I. S. Oliveira, Quantum Inf. Process. \textbf{9}, 757 (2010).

\bibitem{bose} R. Auccaise, J. Teles, T. J. Bonagamba, I. S. Oliveira, E. R. deAzevedo and R. S. Sarthour, J. Chem. Phys. \textbf{130}, 144501 (2009).  

\bibitem{fortunato} E. M. Fortunato, M. A. Pravia, N. Boulant, G. Teklemariam, T. F. Havel and D. G. Cory, J. Chem. Phys. \textbf{116}, 7599 (2002). 



\bibitem{Varsh} D. Varshalovich, A. Moskalev and V. Khersonskii, \emph{Quantum Theory fo Angular Momentum} (World Scientific, Singapore, 1988).


\bibitem{deutsch} D. Deutsch, R. Jozsa, Proc. R. Soc. A \textbf{439}, 553 (1992).


\bibitem{groveralg} L.K. Grover, Physical Review Letters, v. 79, n. 2, p. 325-328. 199


\end{thebibliography}
\end{document}